\documentclass[a4paper,12pt]{article}
\usepackage[cp1251]{inputenc}
\usepackage[english]{babel}
\usepackage[T2A]{fontenc}
\usepackage{amsmath}
\usepackage{amssymb}
\usepackage{graphicx}

\hoffset=+5mm \voffset=+3mm \Roman{table}

\newcounter{tapp}

\begin{document}

\begin{table}[t]
\begin{tabular}{|p{17cm}|}
\hline
\it{Astrophysical Bulletin, 2012, vol. 67, No. 4, pp. 388-398} \\
Translated from Astrofizicheskii Byulleten, 2012, vol. 67,
No. 4, pp. 374-384 \\
\hline
\end{tabular}
\end{table}

\begin{center}
{\Large \bf{PGC~60020: a Polar-Ring Galaxy}}

\vspace{1cm}
     {\bf O.A. Merkulova\footnote{E-mail: olga\_merkulova@list.ru}$^{,\hspace{0.5mm}a}$,
     G.M. Karataeva$^a$, V.A. Yakovleva$^a$, and A.N. Burenkov$^b$}

     { $^a$Astronomical Institute of Saint-Petersburg State University \\
       $^b$Special Astrophysical Observatory, Russian Academy of Sciences}

\abstract{
We present an analysis of new observations of a peculiar galaxy PGC~60020,
obtained with the 6-m BTA telescope of the SAO~RAS with a multimode SCORPIO instrument.
The observational data includes direct images in the B, V, R$_c$ photometric bands and
long-slit spectra in the red range (the H$_\alpha$ line spectral region). Based on
the analysis of these data it was found that PGC60020 belongs to
the type of classical polar-ring galaxies. Its main body is an S0 galaxy,
around the major axis of which a disk of gas, dust and stars is rotating
in the plane inclined at an angle of about~60$^\circ$ to the galactic plane.
A loop-shaped structure stretches from the southern part of this disk
(possibly, a tidal tail) towards the SDSS~J171745.58+404137.1 galaxy.

}
\end{center}

\section{Introduction}

Polar-ring galaxies (PRG) are a rare class of dynamically peculiar systems,
where a ring or a disk consisting of gas, dust and stars is rotating
around the major axis of the main body near the polar plane [1]. It is believed
that such polar structures can emerge either from the interaction or even a merger
of galaxies, or as a result of accretion of gaseous filaments from
the intergalactic medium to the main galaxy ([2–-4],~etc.). The unique geometry of
PRGs allows to obtain the data on the three-dimensional distribution of
the central galaxy’s potential and about the dark halo [5–-7], which only
increases the interest to these objects.

The catalog of PRGs and related objects by Whitmore et al. [1], based on
the photographic images includes 157~galaxies. To date, the existence of
kinematic systems, rotating in different planes (classical~PRGs) is
confirmed only in about two dozen of them.
A small number of known objects does not yet allow to make more or less
definite conclusions about their nature, evolution, and the properties
of the dark halo. Therefore, the search for answers to many issues
related to the emergence, stability and age of polar
rings (PR) is still pertinent. More comprehensive
and accurate data on the kinematics of the stellar
and gaseous components, the properties of the stellar
population and interstellar medium, the processes of
star formation is needed. The discovery of new objects
of this class and their detailed study is of great interest
indeed.

An important step in this direction was the publication
of a new catalog of PRG candidates by Moiseev et al.~[8] in~2011.
This catalog is created based on the results of the Galaxy Zoo
project\footnote{http://www.galaxyzoo.org/}, where the
volunteers performed a visual classification of nearly
a million galaxies from the Sloan Digital Sky Survey
(SDSS\footnote{http://www.sdss.org/}). Based on the preliminary
classification of
the Galaxy Zoo and visual inspections of more than
40~000 images from the Sloan survey, Moiseev and his
colleagues have selected 275 galaxies and compiled
their own Sloan-Based Polar Ring Catalog~(SPRC).
Using the spectral data, obtained both by the authors
of the SPRC catalog and other groups of researchers
(see the references in~[8]), the existence of kinematic
systems, rotating in different planes has already been
confirmed for about~10 galaxies from this catalog.

This paper is devoted to the spectral and photometric
investigation of the PGC~60020 galaxy. In
appearance, this galaxy is similar to the PRGs. This
was first noticed by I.~D.~Karachentsev (SAO~RAS).
He suggested that it belongs to the PRG class of objects
and kindly offered to include PGC~60020 in the
program of our research of the PRG candidates. The
first spectral observations of this galaxy, performed at
the 6-m BTA telescope of the SAO~RAS in~2008 have
confirmed the existence of two kinematic gaseous
subsystems, one of which is related to the main body
of the galaxy, and the other one—to the suspected
PR~[9]. Judging on its external features, PGC~60020
was included in the new SPRC catalog under the
number SPRC-67 as a possible PRG candidate.
We present the results of our observations of
PGC~60020 and their discussion in this paper. The
next section gives a brief about the instruments the
observations were made with, and the data processing
method. Further, sections~3 and~4 present the results
of our study of the morphology and kinematics of
PGC~60020. The conclusion discusses all the results
we obtained.

\section{Observations and processing}

The observations of the PGC~60020 galaxy were
performed at the 6-m BTA telescope of the Special
Astrophysical Observatory of the Russian Academy
of Sciences (SAO~RAS). The \mbox{EEV~42-40~CCD}, sized
2048 $\times$ 2048 pixels (after the 2 $\times$ 2 averaging the pixel
size amounted to \mbox{0.$''$357 $\times$ 0.$''$357}) was used as a radiation
detector.

The photometric observations of the galaxy in
the Johnson B and V bands and Cousins R$_c$-band
were made with the SCORPIO multi-mode focal
reducer~[10], mounted at the primary focus between
May~20 and~21,~2010. Standard stars from the list of
Landolt~[11] were observed for calibration overnight.
The data on the photometric observations are given
in Table~\ref{t:zhurnal_phot}.
\begin{table}[h]
\begin{center}
\caption{Photometric observational data for PGC~60020}
\label{t:zhurnal_phot}
\bigskip
\begin{tabular}{c|c|c|c}
\hline
 Band & Exposure time,    & Seeing, & z, \\
      & frames $\times$ s & arcsec  & deg \\
\hline
 B & 4 $\times$ 350 & 2 & 10--14 \\
 V & 5 $\times$ 180 & 2 & 15--17 \\
 R$_c$ & 11 $\times$ 60 & 1.8--2.2 & 7--9 \\
\hline
\end{tabular}
\end{center}
\end{table}
The observational data reduction was
carried out using the ESO~MIDAS software package.
Correcting for the atmosphere, we used the mean
transparency coefficients of the BTA location~[12].
The accuracy of measurement of the integral magnitude
of the galaxy amounts to~$\pm$0.$^m$1.

Spectral observations were also carried out in the
primary focus of the 6-m telescope with the SCORPIO
focal reducer in the long slit spectral observations
mode, using the VPHG1200R grism; the slit
width amounted to 1$''$, the scale along the slit was
0.$''$357/px, and spectral resolution amounted to 5\AA.
The log of observations is presented in Table~\ref{t:zhurnal_spectr}.
\begin{table}[h]
\begin{center}
\caption{Spectroscopic observational data for PGC~60020}
\label{t:zhurnal_spectr} \bigskip
\begin{tabular}{c|c|c|c|c}
\hline
Date & Exposure time,    & Seeing, & Spectral & PA, \\
     & frames $\times$ s & arcsec  & region, \AA & deg \\
\hline
 27.07.2008 & 2$\times$1200 & 1.6 & 5700--7400 & 115 \\
 27.07.2008 & 2$\times$1200 & 2.4--1.6 & 5700--7400 & --10 \\
 19.07.2010 & 5$\times$1200 & 1.7 & 5700--7400 & 115 \\
 20.07.2010 & 8$\times$1200 & 1.2 & 5700--7400 & --10 \\[1mm]
\hline
\end{tabular}
\end{center}
\end{table}

The observations in the long slit mode were
performed in~2008 and~2010 in the red spectral
range, containing the following emission lines: H$_\beta$,
[NII] $\lambda\lambda$ 6548, 6584\AA. The reduction of the obtained
data was carried out using the standard procedures
within the ESO~MIDAS environment. After the
primary reduction, to increase the signal-to-noise
ratio we summed all the obtained spectra, and the
resulting spectrum was smoothed along the slit by a
3-pixel-high rectangular window. The radial velocities
of the gas component were measured by the positions
of the centers of Gaussians, inscribed into the
emission lines. The accuracy of these measurements
was estimated by the night sky [OI] $\lambda$ 6300\AA~line
and amounted to 10--15~km/s. The cross-correlation
method~[13] was applied to determine the radial
velocity from the absorption lines.

\section{Multicolor photometry results}

\subsection{Features of the photometric structure
in PGC~60020}

\begin{figure}[h!]
    \vspace*{-0.0cm}
    \hspace*{-0.0cm}
    \vbox{ \includegraphics{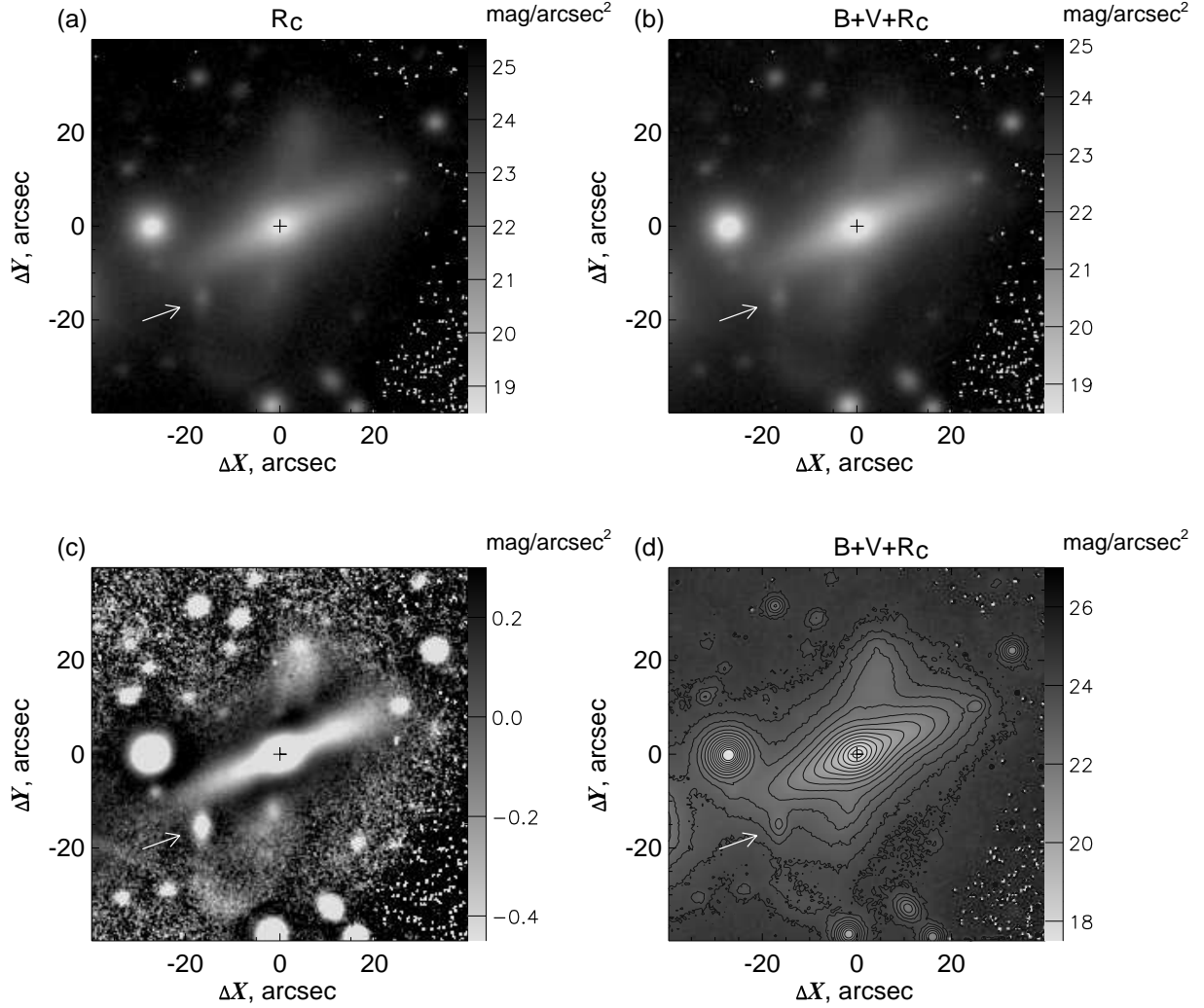}
             } \par
\vspace*{13.cm} \hspace*{-0.0cm}
\caption{PGC~60020 and its companion galaxy SDSS~J171745.58+404137.1: (a) the image in the R$_c$-band; (b) the total image
in the B, V, R$_c$-bands; (c) the residual image in the R$_c$-band,
obtained by subtracting the image, processed by a median filter;
(d) the total image in the B, V, R$_c$-bands with the isophote 
contours. N is on top, E is on the left. An arrow points to 
the SDSS~J171745.58+404137.1 galaxy.}
\label{f:f1}
\end{figure}

The photometric observations of PGC~60020 were
obtained in three filters, B, V, R$_c$. The image of the
galaxy in the R$_c$-band is given in Fig.~\ref{f:f1}a.
Its main
body has a lenticular shape: a bright ellipsoidal central
region and a faint external disk, visible at an
angle to the sky plane. By appearance, this galaxy
can be attributed to the S0 morphological type, although
the LEDA database designates it as an elliptical
with a question mark (E?). To the north and
south of the galactic plane along the --10$^\circ$ direction,
extended luminous regions are visible, the presence of
which makes PGC~60020 very similar to such classic
PRGs as NGC~2685, ESO~415-G26, IC~1689, and
AM~2020-504~[4,~14--16]. A comparison of direct
images obtained at the 6-m telescope with the SDSS
images has shown that the BTA images are deeper,
hence only our data were used in the analysis of the
external structure of the galaxy.

The outer regions of the main body and the suspected
PR are very faint. We summed the frames,
obtained in three filters, the total image is shown in
Figs.~\ref{f:f1}b and d, where the features of the outer regions
of PGC~60020 and its immediate surroundings are
more discernible. In the vicinity of PGC~60020 (in a
circle of a 1$'$ radius) three faint galaxies are observed.
Visually inspecting the total B+V+R$_c$ image with
the superimposed isophotes (Fig.~\ref{f:f1}d) one gets an
impression that from the southern end of the suspected
PR a luminous arc stretches to one of these
objects, located at a distance of 22.$''$2 to the south-east
from the PGC~60020 center. This object is identified
as the SDSS~J171745.58+404137.1 galaxy. However,
to date, neither its morphology, nor redshift are
known. For the other two faint galaxies such data are
also absent. The presence of a luminous arc between
PGC~60020 and SDSS~J171745.58+404137.1 allows
to suspect that the latter belongs to the close
neighborhood of PGC~60020, and further in this paper
we shall call the given object a ``companion'' galaxy.

To clarify the structural features of PGC~60020,
the unsharp masking image manipulation technique
was applied, where the Gaussian smoothing was performed
with a $\sigma \approx$ 7$''$ window of the studied galaxy.
Figure~\ref{f:f1}c presents a mask, obtained after subtracting
the smoothed image in the R$_c$ filter from the initial
image. It is clear that the main body of the galaxy
consists of a bright bulge and an inclined disk, and
that the diameters of the major axes of the galaxy and
the suspected ring are about the same (the outer parts
of the ring extend up to nearly 20$''$ from the center).
The companion galaxy is roughly elliptical in shape.
In this image, the luminous arc, mentioned above is
clearly discernible.

\begin{figure}[h!]
    \vspace*{-0.0cm}
    \hspace*{-0.0cm}
    \vbox{ \includegraphics{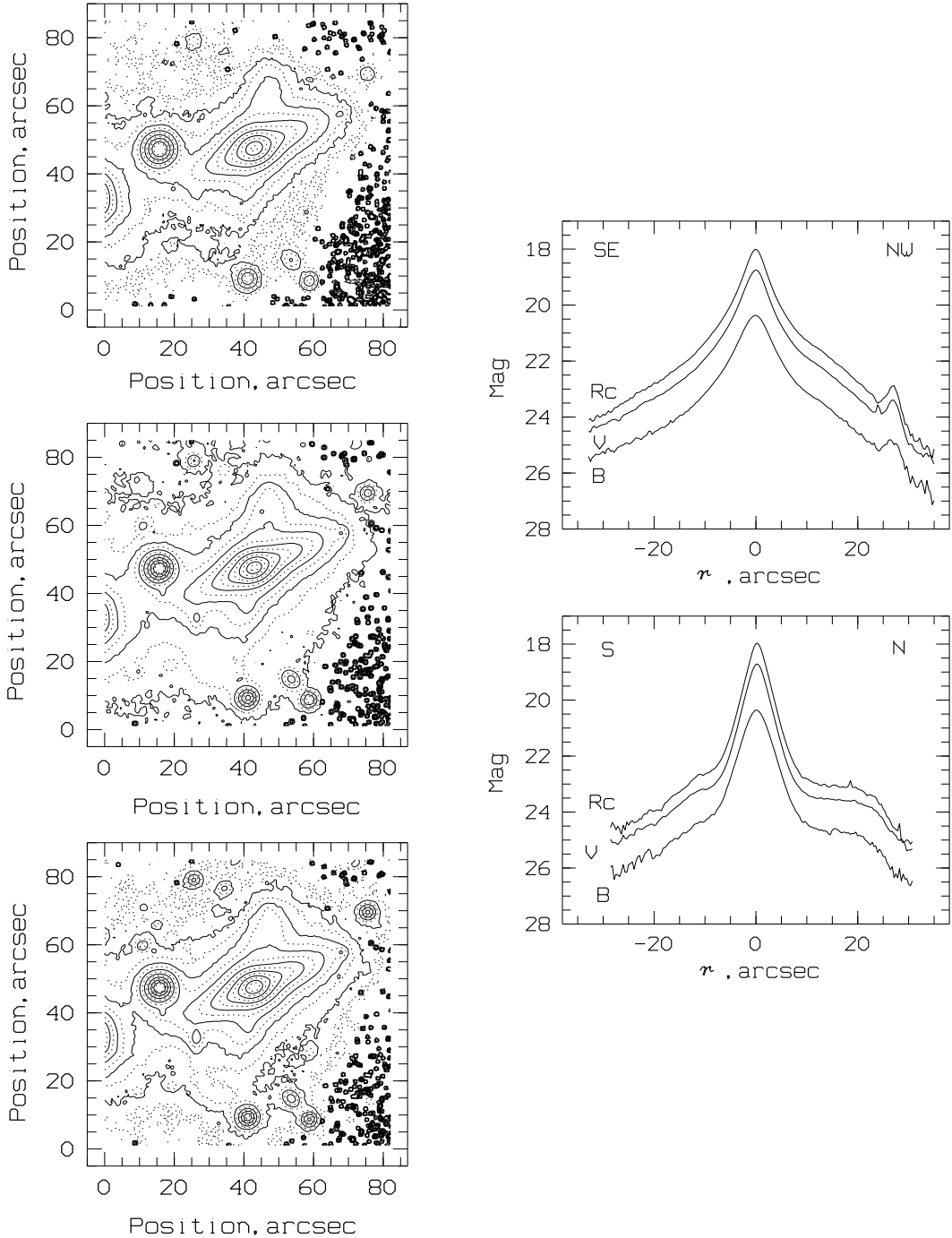}
             } \par
\vspace*{17.cm} \hspace*{-0.0cm}
\caption{PGC~60020: left column (from top to bottom): 
the galaxy isophotes in the B, V, R$_c$-bands (in increments 
equal to 0.$^m$5/$\Box''$; the outer isophote in the B-band 
corresponds to the surface brightness of 26.$^m$5, in the 
V-band---25.$^m$5, and in the R$_c$-band---25.$^m$5; 
N is on top, E is on the left); right column: 
the B, V, R$_c$ image sections along the major axes of 
the main body (top) and the ring (bottom).}
\label{f:f2}
\end{figure}

Figure~\ref{f:f2} presents the isophotes of PGC~60020
in the B, V, R$_c$-bands and the photometric sections
along the major axes of the main body and the ring.
The shape of the isophotes in all colors is about the
same. It should be noted that the presence of the
ring greatly distorts the shape of the outer isophotes
of the main body of PGC~60020. For the analysis
of the photometric structure, we used the technique,
based on the Fourier series expansion of the isophote
deviation from elliptical, suggested in~[17].

\begin{figure}[h!]
    \vspace*{-0.0cm}
    \hspace*{-0.0cm}
    \vbox{ \includegraphics{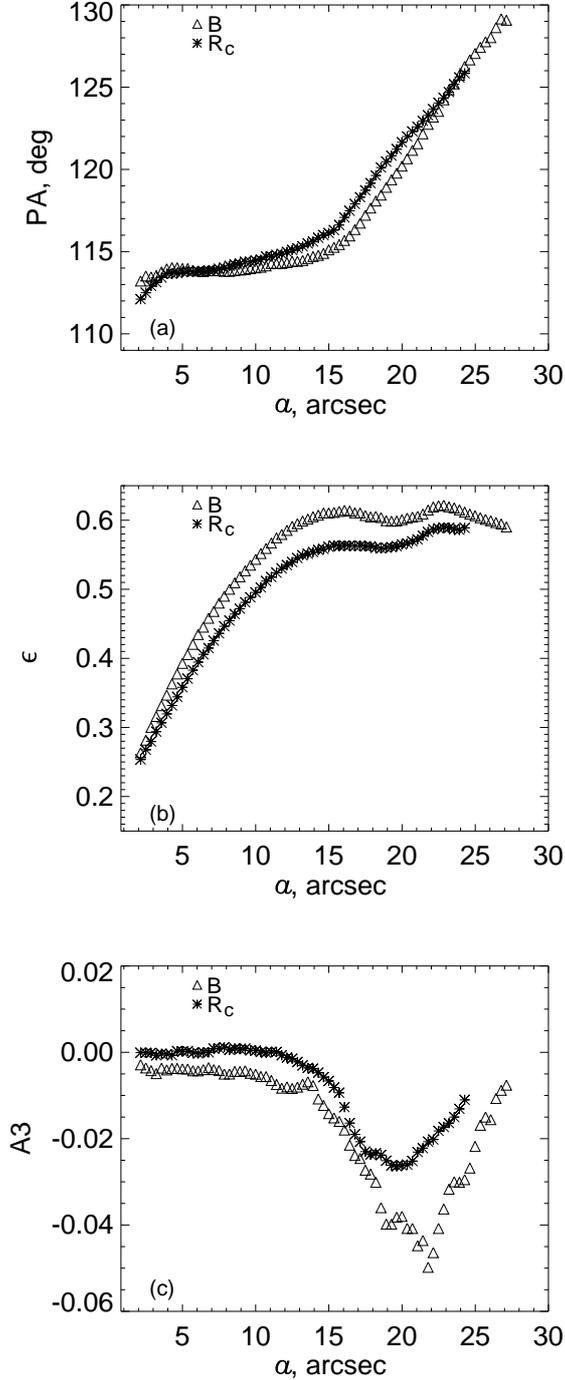}
             } \par
\vspace*{17.5cm} \hspace*{-0.0cm}
\caption{PGC~60020: the characteristics of the isophote
shapes in the B (triangles) and R$_c$ (asterisk) bands, 
depending on the semi-major axis of the ellipse, inscribed
in the isophote: (a) position angle, (b) ellipticity, 
(c) the coefficient of the third harmonic A3.}
\label{f:f3}
\end{figure}

The isophote analysis (of the frames in the B and
R$_c$-bands) was performed in the IRAF environment.
For each value of the major axis ($a$) we computed the
ellipticity ($\epsilon_{ell}$) and the position angle of the major axis
of the inscribed ellipses (PA$_{ell}$), which was measured
from the direction to north towards east, as well as
the dimensionless coefficients of the third and the
fourth harmonics in the Fourier series expansion. The
difference between the coefficients of these harmonics,
especially the third one, in the B and R$_c$-bands
is a reliable indicator of the presence of dust in the
galaxy~[18]. The findings of this analysis are presented
in Fig.~\ref{f:f3}.

Overall, the behavior of the $\epsilon_{ell}$ and PA$_{ell}$ ellipse
parameter variability in both filters is about the same
(Fig.~\ref{f:f3}a and b). In the interval from 2$''$ to 15$''$ the
ellipticity smoothly varies from 0.25 to about 0.56 in
the R$_c$ filter, and up to 0.60 in the B filter, while at
large distances it is approximately constant. In the
galactic region from 4$''$ to 16$''$, PA$_{ell}$ varies little and
on the average PA$_{ell}$ = 115$^\circ$. Further, a variation of
the position angle is observed, and at the distance of
24$''$, PA$_{ell}$ $\approx$ 125$^\circ$. Such a variation of PA$_{ell}$ is most
likely due to the fact that at these distances, the surface
brightness of the ring becomes comparable to the
surface brightness of the main body of PGC~60020.
In this case, the shape of the outer isophotes gets
distorted, and the position angle of the major axis of
the galaxy itself (PA$_{gal}$) does not vary (see Fig.~\ref{f:f2}).

In the central region of the galaxy ($a$ $\leq$ 10--15$''$)
the coefficients of the cosine and sine of the third
and fourth harmonics (A3, A4, B3, B4) are close to
zero in both filters, then their values become nonzero,
but the differences in the B and R$_c$ filters are
observed only in the A3 coefficient. Figure~\ref{f:f3}c shows
that up to $a$ $\approx$ 15$''$ the value of A3 is approximately
equal to zero. Further, it becomes negative, and a little
difference emerges between the filters, amounting to
0.02--0.03. Based on this we can conclude that the
main body of the galaxy is poor in dust, though a small
amount of dust may be present in the ring.

As a result of studying the photometric structure
of PGC~60020, we have taken that the position angle
of the major axis of the galaxy PA$_{gal}$ = 115$^\circ$ $\pm$ 2$^\circ$, and
the angle of inclination of the disk plane to the sky
plane i$_{gal}$ = 65$^\circ$ $\pm$ 2$^\circ$. The ellipticity $\epsilon$ was taken to
be 0.56 (the value obtained in the R$_c$-band), since in
the analysis of the galaxy image in the K$_s$-band from
the Two-Micron Sky Survey (2MASS), the value of
$\epsilon$ proved to be closer to the ellipticity measured in
the R$_c$-band. The position angle of the major axis of
the ring is --10$^\circ$ $\pm$ 2$^\circ$, and the angle of inclination to
the plane of the sky is approximately 79$^\circ$ $\pm$ 5$^\circ$ (the
ellipticity is 0.8). Knowing the inclination of the disk
plane of the main body of the galaxy and the ring to the
sky plane (65$^\circ$ and 79$^\circ$, respectively) and the positions
of their major axes (115$^\circ$ and --10$^\circ$, respectively), we
can find the angle between the disk and the ring from
the expression:
\begin{center}
    cos $\Delta i$ = $\pm$ sin$i_1$ sin$i_2$ cos($PA_1 - PA_2$) + cos$i_1$ cos$i_2$,
\end{center}
where $i_1$, $i_2$ are the disk and ring inclination angles
to the sky plane, PA$_1$, PA$_2$ are the position angles of
the major axes of the galactic disk and the ring. This
angle was found to be 54$^\circ$ $\pm$ 4$^\circ$ and 115$^\circ$ $\pm$ 4$^\circ$.

In conclusion, let us say a few words about the
structure of the companion galaxy. Its surface brightness
is low, the images in all filters are amorphous,
without any features. As noted above, it is elliptical
in shape. The scale of the major axis of the ellipse
in the B-band from the $\mu_B$ = 25.0~mag/$\Box''$ isophote
is about~4.$''$3, the ratio of the semiaxes amounts to
about~0.56, and the ellipticity is approximately~0.4.
This value corresponds to the ellipticity of E4--E5
galaxies.

\subsection{Photometric properties of galaxies}

Apparent integral magnitudes and color indices
of the PGC~60020 galaxy and its companion were
determined using the multiaperture photometry~[19]
with the accuracy of $\pm$0.$^m$1, as stated above. The
integral values, corrected for the extinction in our
Galaxy~[20] are listed in Table~\ref{t:data}. 
\begin{table}[h]
\begin{center}
\caption{Main characteristics of PGC~60020 and 
the suspected companion galaxy}
\bigskip
\begin{tabular}{c|c|c}
\hline
Characteristics & PGC~60020 & Companion \\
\hline
Morphological type & E?$^*$ & -- \\
                   & S0     & E4-E5 \\
Distance ($H_0$ = 72~km/s/Mpc) & 118~Mpc & -- \\
Scale & 0.57~kpc in 1$''$ & -- \\
$V_{gal}$, km/s & 8507 $\pm$ 60$^{**}$ & -- \\
                                 & 8489 $\pm$ 20 & -- \\
Semi-major axis $a$ ($\mu_B$ = 25.0) & 23.8$''$ (13.6~kpc) & 4.3$''$ \\
Position angle of the major axis $PA_{gal}$ & 115$^\circ$ & 0$^\circ$ \\
Inclination angle $i_{gal}$ & 65$^\circ$ & -- \\
Semi-major axis of the ring $a_{ring}$ ($\mu_B$ = 25.0) & 21.4$''$ (12.2~Kpc) & -- \\
Position angle of the ring's major axis $PA_{ring}$ & --10$^\circ$ & -- \\
Inclination angle of the ring $i_{ring}$ & 79$^\circ$ & -- \\
B$_{t,0}$,~mag & 15.6 & 17.7 \\
(B--V)$_0$,~mag & 1.15 & 1.06 \\
(V--R)$_0$,~mag & 0.58 & 0.47 \\
$M_B$,~mag & --19.9 & -- \\
B/D & 0.6 & -- \\[1mm]
\hline
\end{tabular}

\vskip0.3cm
{\footnotesize * The data adopted from the HyperLeda database
(http://leda.univ-lyon1.fr/),\\
** The data adopted from the NED database (http://ned.ipac.caltech.edu)}
\label{t:data}
\end{center}
\end{table}
The apparent magnitude
of PGC~60020 that we obtained, $B_{t,0}$ = 15.$^m$6
coincides with the value given in the LEDA database
within the limits of error. Calculating the absolute
magnitude $M_B$, we have introduced a correction for
redshift. The integral B-V color index of PGC~60020
is by about 0.$^m$2 redder than the average color index of
the early-type galaxies. As shown above, this fact can
not be explained by the presence of dust. Regarding
the value of V--R$_c$, we can assume that within the
errors its value coincides with the color indices of
E--S0-type galaxies. Since a major part of the ring
is projected on the main body of the galaxy, its luminosity
can not be reliably estimated. We can only
find the apparent integral magnitude of two prominent
luminous regions located to the north and south of the
main body of the galaxy. Their integral magnitude is
about 19$^m$, which is about 4\% of the total luminosity
of PGC~60020.

Integral color indices of the companion are similar
to the indices of elliptical galaxies within the limits of
error.

\begin{figure}[h!]
    \vspace*{-0.0cm}
    \hspace*{-0.0cm}
    \vbox{ \includegraphics{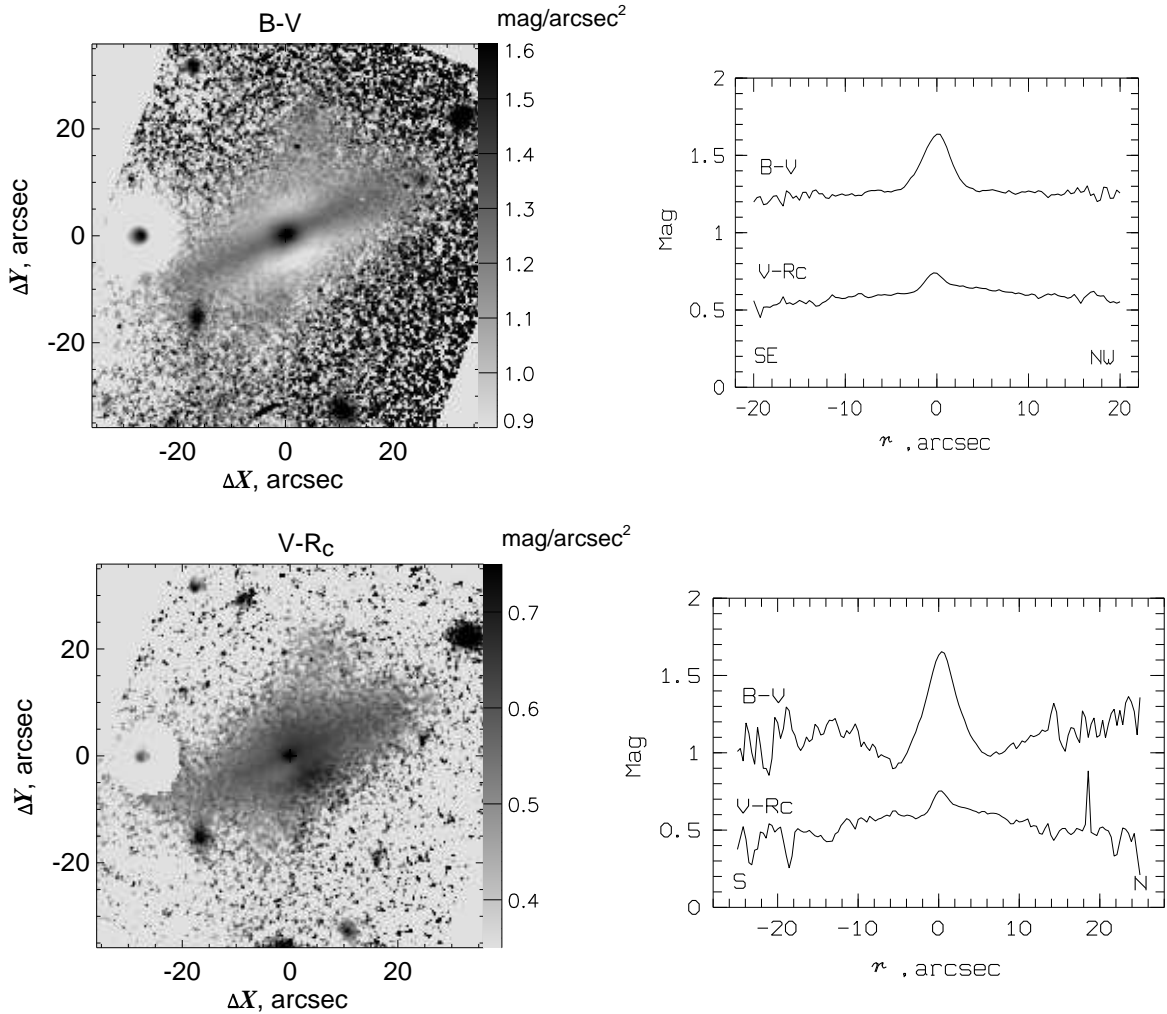}
             } \par
\vspace*{12.cm} \hspace*{-0.0cm}
\caption{PGC~60020: left column: the B--V (top) and
V--R$_c$ (bottom) distributions (N is on top, E is on the left); 
right column: the B--V and V--R$_c$ sections along 
the major axes of the main body (top) and the ring (bottom).}
\label{f:f4}
\end{figure}

The distributions of B--V and V--R$_c$ color indices,
as well as the B--V and V--R$_c$ sections along the
major axes of the main body and the ring are demonstrated
in Fig.~\ref{f:f4}.
The reddest color indices B--V
are observed in the circumnuclear region ($r$ < 2--3$''$)
of PGC~60020 and in the center of the companion
galaxy, where B--V reaches 1.$^m$6 and about 1.$^m$5, respectively.
In the plane of the disk of PGC~60020,
the value of B--V is constant within the error limits
and amounts to \mbox{1.$^m$2 $\div$ 1.$^m$3}. The color index decreases
with distance from the galactic nucleus, while in the
direction to NE and SW at the distance of approximately
4--5$''$ two regions with \mbox{B--V = 0.$^m$8 $\div$ 0.$^m$9} are
observed, of which the SW region is more extended.
It extends approximately in parallel to the galactic
plane, and its size in this direction is about 12--14$''$ .

The distribution of the V--R$_c$ color index is more
uniform, and the value of V--R$_c$ changes little: from
about 0.$^m$7 in the nucleus of PGC~60020 to 0.$^m$5 on the
periphery. As for the companion, its V--R$_c$ color stays
almost constant and is equal to about 0.$^m$7.

In the region of the ring, the color indices bluer
than in the main body of the galaxy are observed
(B--V $\approx$ 0.$^m$9--1.$^m$0). A considerable scatter of B--V
values at the section along the major axis of the ring
(Fig.~\ref{f:f4}, bottom right plot) at $r$ > 16$''$ is explained by
a clumpy structure of the ring and a deterioration in
accuracy due to the low surface brightness. At the
distance of approximately 5--6$''$ to the north and south
from the center, where the given section crosses two
blue regions, whose existence was mentioned above,
a decrease of B--V is observed, while from the south
this effect is more noticeable. In general, it should be
noted that the color indices of the ring are by about
0.$^m$2--0.$^m$3 higher than the color indices of the rings in
classic PRGs (see, e.g.,~[4]).

\begin{figure}[h!]
    \vspace*{-0.0cm}
    \hspace*{-0.0cm}
    \vbox{ \includegraphics{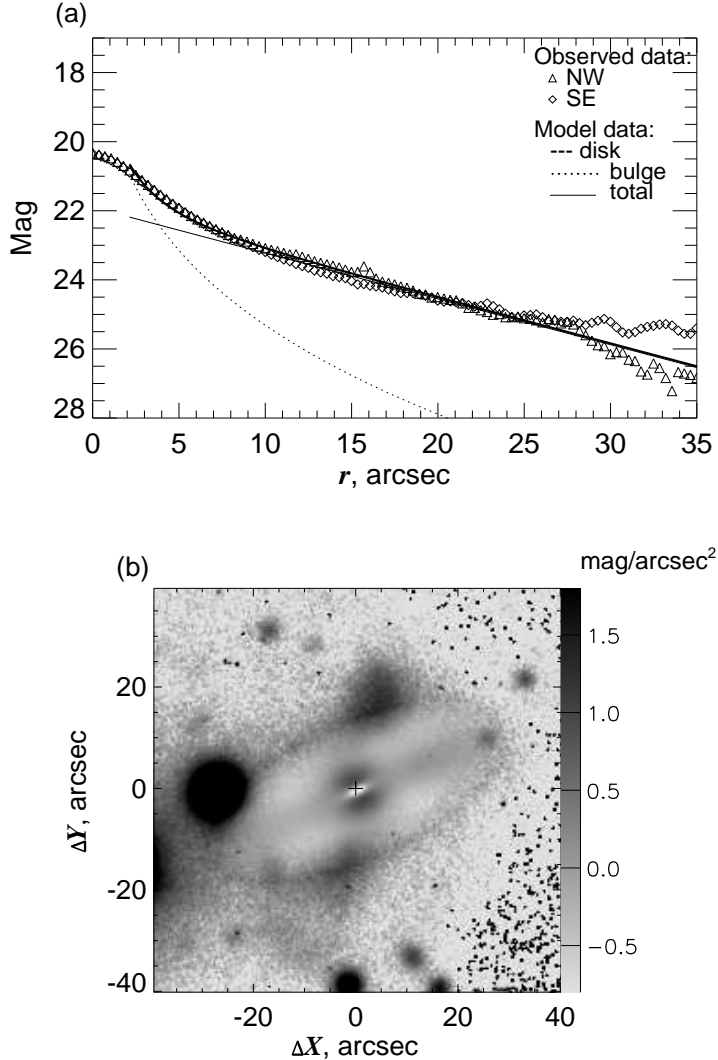}
             } \par
\vspace*{13.5cm} \hspace*{-0.0cm}
\caption{PGC~60020: (a) the observed brightness profile
along the major axis with \mbox{PA = 115$^\circ$} and 
its decomposition into components; (b) residual image 
after subtracting the two-dimensional model from 
the B-band image.}
\label{f:f5}
\end{figure}

Figure~\ref{f:f2} (right column, at the top) shows the
sections along the major axis of the galaxy in three
filters. For the decomposition into components, we
used the B-band section along the major axis. We
can isolate two areas in this section: the central area
(up to $r$ $\approx$ 7$''$ from the center) with a steeper profile
and an extended external area (10$''$ $\leq$ $r$ $\leq$ 25$''$), where
the profile is less steep. The brightness profile in the
outer region almost throughout its length is well represented
by an exponential law with a scale factor
$h_d$ = 8.$''$2 (4.7~kpc) and central surface brightness
$\mu_d$ = 21.9~mag/$\Box''$ (Fig.~\ref{f:f5}a).
The central structure
(the bulge) can be described by the de Vaucouleurs
law with an effective radius $R_{e,b}$ = 1.$''$4 (0.8~kpc) and
effective surface brightness $\mu_{e,b}$ = 20.03~mag/$\Box''$.

Using the found parameters of the bulge and the
disk a two-dimensional model of the galaxy was constructed.
At that, we assumed that the position angle
of the photometric axis amounts to 115$^\circ$, and the
inclination of the bulge and the disk of PGC~60020
to the sky plane is 65$^\circ$. The residual brightness distribution,
obtained by subtracting the two-dimensional
model from the galaxy image in the B-band is shown
in Fig.~\ref{f:f5}b. It clearly demonstrates an increase of differences
between the observed and model brightness
distributions in the region of the ring.

We also estimated the total luminosity ratio of the
bulge and the disk B/D. It was found to be 0.6, a value
typical of S0 or Sa galaxies~[21].

\section{Results of spectroscopic observations}

Long-silt spectra of PGC~60020 were obtained
along the major axes of the main body \mbox{(PA = 115$^\circ$)}
and suspected ring (PA = --10$^\circ$). Both spectra show
a high-energy continuum with strong absorption
lines of the NaI~D~$\lambda$5890, 5896~\AA~doublet and
the blend of FeI+CaI+BaII lines around $\lambda$6495~\AA.
Among the emission lines, the [NII] and [SII] doublets
are present, as well as the H$_\alpha$ line. However,
in our spectra the [SII] lines of the object fall on
the lines of the sky, the line of H$_\alpha$ in the emission
is superimposed over a strong absorption line, and
the [NII]~6548~\AA~line can be reliably identified only
in the nuclear region ($r$ < 2$''$), sinking in the noise at
large distances from the center. Therefore, the radial
velocity curves were built based on the [NII]~$\lambda$6584~\AA~line.

\begin{figure}[h!]
    \vspace*{-0.0cm}
    \hspace*{-0.0cm}
    \vbox{ \includegraphics{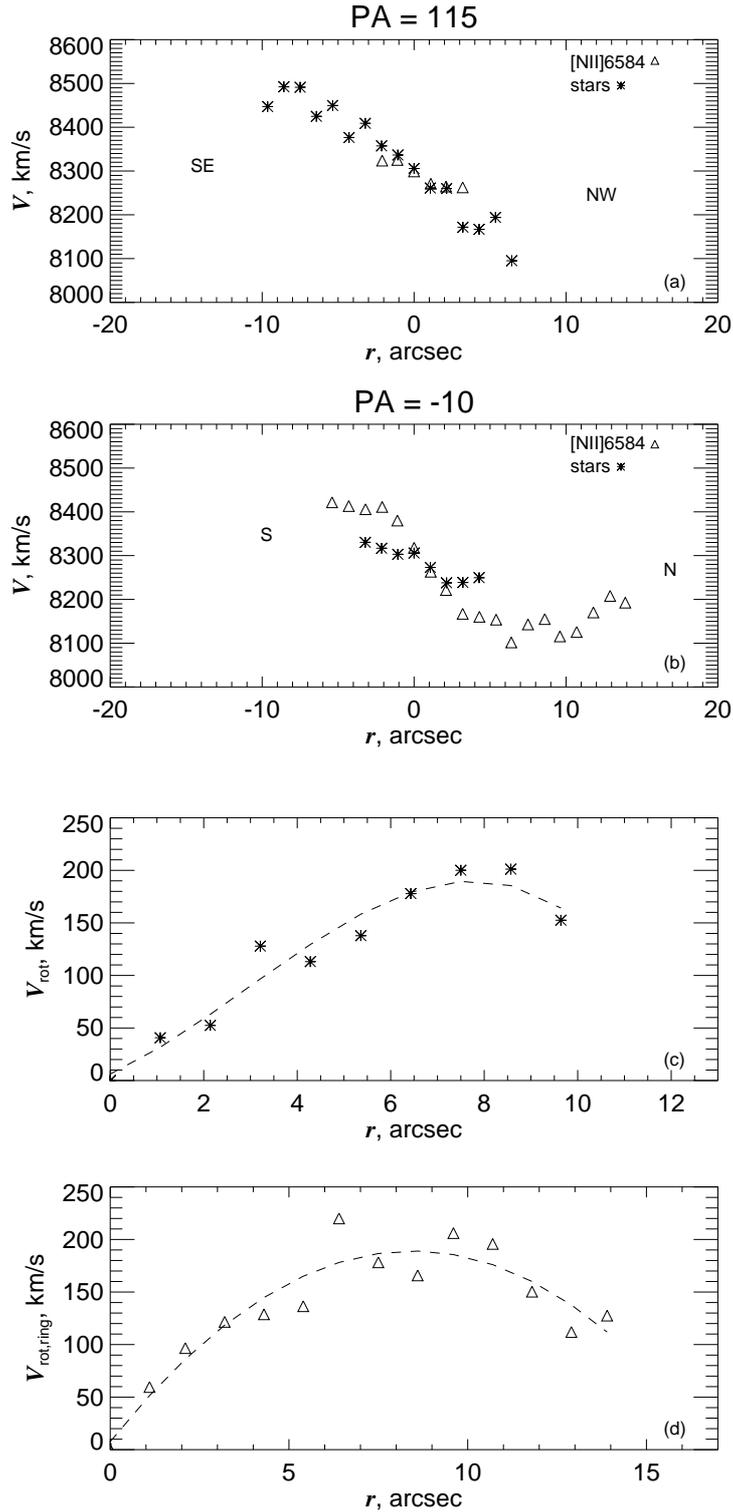}
             } \par
\vspace*{19.5cm} \hspace*{-0.0cm}
\caption{PGC~60020: radial velocity curves of ionized gas
and stars along the major axes of the main body (a) and
the suspected ring (b); rotation curves of the main body
stellar component (c) and ionized gas of the ring (d).
The signs (the asterisks and triangles, respectively) mark
the observed rotation curves, and the dashed lines---the
average smoothed rotation curves.}
\label{f:f6}
\end{figure}

To construct the stellar radial velocity curves using
the cross-correlation method, the spectrum of
the galactic nucleus was used, since the spectra of
template stars were not be obtained. We succeeded
in constructing the curves of stellar radial velocities
along the major axis of the galaxy (PA = 115$^\circ$), and
along PA = --10$^\circ$ up to the distances of $r$ = 8--10$''$,
and about 4$''$ from the center, respectively (Fig.~\ref{f:f6}).
These curves do not manifest any specific features,
they are typical of galactic disks: the maximum velocity
gradient along the major axis of the disk and
a small velocity gradient along the direction between
the major and minor axes (PA = --10$^\circ$). Assuming
that the stars rotate in circular orbits, and given the
inclination of the main body to the sky plane (65$^\circ$),
we recalculated the curve of stellar radial velocities
of stars along \mbox{PA = --10$^\circ$} to the curve along the
major axis. Both curves coincided within the error of
observations. This means that the absorption lines in
the spectra along PA = 115$^\circ$ and --10$^\circ$ belong to the
stellar the population of the main body of the galaxy.
If we assume that the photometric axis of the main
body coincides with the dynamic axis, we can then
construct in the region up to $r$ $\approx$ 10$''$ the rotation
curve of the galaxy (Fig.~\ref{f:f6}c). The velocity gradient
in the rectilinear section is about 44~km/s/kpc.
It is possible that at a distance of $R_{max}$ = 8$''$--10$''$
(4.6--5.7~kpc) the rotation curve reaches a plateau,
and then the maximum velocity amounts to approximately
200~km/s. The estimate of the mass, enclosed
within the radius of $R_{max}$ yields the value of
$M(R_{max})$ $\geq$ 4.3 $\times$ 10$^{10}$ $M_\odot$.

The radial velocity curves of ionized gas are shown
in Fig.~\ref{f:f6}. Note that in the circumnuclear region of the
galaxy the full width at half maximum (FWHM) of
the [NII]~$\lambda$6584~\AA~line is about 460~km/s. Along the
major axis of the main body, this line is visible only in
the circumnuclear region ($r$ $\leq$ 3--4$''$). A similar pattern
is observed in some classical PRGs, for example,
in NGC~2685~[22]. The radial velocity gradient along
the major axis is about 33~km/s/kpc (Fig.~\ref{f:f6}a). Since
the central region of the galaxy does not reveal any
features, we can assume that the photometric and dynamical
centers coincide, and the heliocentric velocity
of the galactic center is about 8308 $\pm$ 20~km/s. The
radial velocity gradient of ionized gas along the major
axis of the galaxy proved to be slightly lower than the
velocity gradient in the rectilinear region of the stellar
radial velocity curve.

The line of nitrogen can be traced from the northern
part to 14$''$ from the center, and from the southern
part only to 7$''$ along the major axis of the ring. This
result is not surprising, since the northern part of the
ring is much brighter than its southern counterpart
(see Fig.~\ref{f:f1}a). The radial velocity curve along the
major axis of the ring demonstrates that the gas of
the disk is rotating around the major axis of the main
body of the galaxy. In the central region with the
\mbox{$r$ $\approx$ 3$''$} radius there is a straight section with a velocity
gradient of approximately 81~km/s/kpc. Farther off,
the velocity gradient declines, and in the northern
part the radial velocity curve reaches a plateau. At a
distance of $r$ $\approx$ 8$''$ the peak radial velocity is reached
with respect to the system center, amounting to approximately
--170~km/s, then it decreases to come up
to about --120~km/s at $r$ $\approx$ 13$''$.

A large velocity gradient in the rectilinear region of
the curve indicates that the polar structure, observed
in PGC~60020 is evidently a \textit{polar disk}, and the
emission of ionized gas, belonging to this disk, makes
a significant contribution to the total emission of gas
in the central region of the galaxy. This also explains a
smaller radial velocity gradient of ionized gas relative
to the gradient of radial velocity of stars along the
major axis of the galaxy (Fig.~\ref{f:f6}a). Further in the text
we shall use the term ``polar disk'' instead of the ``polar
ring''. Figure~\ref{f:f6}d shows the observed rotation curve of
ionized gas in the polar disk, constructed in the assumption
that the gas is rotating in circular orbits, the
position angle of the disk's major axis is --10$^\circ$, and its
angle of inclination to the sky plane is 79$^\circ$. The peak
velocity is reached at about the same distances from
the center, as those revealed in the rotation curve of
the stellar component of the main body of the galaxy,
and is about 190~km/s.

We have also attempted to obtain the spectrum
of the companion galaxy (the observing set of
May~15 to~16,~2012) in order to determine its redshift.
However, the object is faint and due to bad weather
conditions the obtained spectrum has revealed the
stellar continuum only, resulting in the failure to estimate
the redshift of SDSS~J171745.58+404137.1.

\section{Discussion and conclusions}

Before the final conclusions, let us discuss some
features of the PGC~60020 galaxy, discovered in the
course of our research. The characteristics we have
determined are presented in Table~\ref{t:data}.

It was earlier noted that in some aspects the main
body of the galaxy can be attributed to the S0 type,
however, its integral color indices proved to be redder
than those typical of the elliptical galaxies, and even
more so than in the classical PRGs with S0-type
main bodies. The analysis of the isophote shapes in
the B and R$_c$-bands did not reveal any presence of
dust in the main body of the galaxy. Consider other
possible explanations of this feature.

One reason may be due to the fact that PGC~60020
is poor in gas. This is confirmed by the presence of
emission lines only in its central region
($r$ $\leq$ 3--4$''$),
where the total emission of ionized gas of the polar
and galactic disks is observed. Moreover, the outer
parts of the galactic disk do not reveal any noticeable
decrease in the color indices, typical of the majority of
classical PRGs.

On the other hand, the redder color indices may
be possibly related with the features of stellar population
in the circumnuclear region of the galaxy. Large
B--V color indices, reaching 1.$^m$6 in this region, and
a powerful continuum with strong absorption lines
imply the presence of a large number of late-type
stars. However, we are currently not in possession
of the data required to perform the study of stellar
populations (age, metallicity). We consider unreliable
the use of color indices in the case of such a peculiar
galaxy as PGC~60020 for such an analysis. To make
the final conclusions, high quality spectra with a good
spectral resolution are required for the absorption line
analysis.

Two areas with bluer indices are visible in the
distribution of B--V color indices (see
Fig.~\ref{f:f4}, top left
plot) towards the NE and SW of the galactic center at
a distance of about 4--5$''$. Blue color indices are most
likely related to the fact that the polar disk radiation
makes a noticeable contribution in these regions.

Based on the analysis of the photometric and spectral
data we have obtained, we can make the following
conclusions about the structure and nature of the
PGC~60020 galaxy.

1. PGC~60020 is a classical polar-ring galaxy. This
is revealed by the features of its photometric structure,
and, mainly, the presence of two kinematic subsystems,
rotating in different planes. The diameter of the
polar disk is not greater than the diameter of the main
body of the galaxy, hence PGC~60020 belongs to the
group of PRGs with the so-called inner polar rings
(NGC~2685, IC~1689, AM~2020-504,~etc.).

2. The main body of PGC~60020 is an S0 galaxy.
This is confirmed by the analysis of the isophote
shapes, and by decomposing the brightness profile
into the components along the major axis (bulge +
disk). The photometric characteristics of the main
body, such as the absolute magnitude ($M_B$ = --19.$^m$9),
the parameters of the bulge and the disk ($h_d$ = 8.$''$2
(4.7~kpc), $\mu_d$ = 21.9 mag/arcsec$^2$; $R_{e,b}$ =
1.$''$4
(0.8~kpc), $\mu_{e,b}$ = 20.03 mag/arcsec$^2$),
and the bulge-to-disk ratio (0.6) are typical of
the S0 galaxies
with polar rings~[4,~23]. The estimate of the mass
of the galaxy ($\geq$ 4.3 $\times$ 10$^{10}$ $M_\odot$)
and the mass-to-luminosity
ratio ($\geq$ 6.8 $M_\odot/L_\odot$) are also consistent
with the above assertion.

3. The presence of emission lines ([NII] and H$_\alpha$)
in the circumnuclear region of PGC~60020 suggests
the ongoing activity of the galactic nuclei. The fact
that the forbidden nitrogen line in this region is much
brighter than the H$_\alpha$ line is indicative of the gas
emission from the shock ionization. In addition, the
nitrogen line FWHM is about 460~km/s. The above
data allow us to make an assumption that the nucleus
of the PGC~60020 galaxy has the LINER? characteristics
(see, e.g.,~[24]). However, the spectral observations
in the blue region are yet required for the final
conclusion.

4. Around the main body of the galaxy, in the
plane inclined to the galactic plane by an angle of
about 60$^\circ$ (our estimates), a disk, consisting of gas,
stars and dust is rotating. The disk structure is very
heterogeneous. The plane of the disk is inclined to the
polar plane of the galaxy by an angle of about 30$^\circ$.

5. A loop-like structure, possibly a tidal tail
stretches from the southern part of the polar disk
to the companion galaxy. However, it is currently
unknown whether these galaxies do interact or we are
simply observing a projection effect. If these galaxies
are indeed close and interacting, the polar disk could
be formed as a result of accretion of matter from the
companion galaxy to PGC~60020.

6. The SDSS~J171745.58+404137.1 galaxy is
classified as E4--E5; red color indices and the absence
of emission lines in its spectrum do not contradict this
classification.

\bigskip

{\large \bf Acknowledgements}

The authors are grateful to the Committee of
Large Telescopes for allocating the observing time
at the 6-m BTA telescope, and the staff of the SAO~RAS, 
namely A.~V.~Moiseev for his help in the observations
with the 6-m telescope, and I.~D.~Karachentsev
for drawing our attention to the PGC~60020
galaxy as a possible PRG candidate. The initial work
was supported by the Russian Foundation for Basic
Research (grant no.~05-02-17548). O.~A.~Merkulova
is also grateful to the support from the Federal target
program Research and Pedagogical Cadre for Innovative
Russia (no.~12.740.11.0133). The observations
at the 6-m BTA telescope are performed with the
financial support of the Ministry of Education and
Science of the Russian Federation (state contracts
no.~16.552.11.7028,~16.518.11.7073).

\bigskip

{\large \bf References}
\begin{flushleft}

1. B. C. Whitmore, R. A. Lucas, D. B. McElroy, et al.,
Astronom. J. $\textbf{100}$, 1489 (1990).

2. K. Bekki, Astrophys. J. $\textbf{499}$, 635 (1998).

3. F. Bournaud and F. Combes, Astronom. and Astrophys.
$\textbf{401}$, 817 (2003).

4. V. P. Reshetnikov and N. Ya. Sotnikova, Astronom.
and Astrophys. $\textbf{325}$, 933 (1997).

5. E. Iodice, M. Arnaboldi, R. P. Saglia, et al., Astrophys.
J. $\textbf{643}$, 200 (2006).

6. P. D. Sackett, H. Rix, B. J. Jarvis, and K. C. Freeman,
Astrophys. J. $\textbf{436}$, 629 (1994).

7. F. Schweizer, B. C. Whitmore, and V. C. Ruben, Astronom.
J. $\textbf{88}$, 909 (1983).

8. A. V. Moiseev, K. I. Smirnova, A. A. Smirnova, and
V. P. Reshetnikov, Monthly Notices Roy. Astronom.
Soc. $\textbf{418}$, 244 (2011).

9. G. M. Karataeva, O. A. Merkulova, A. N. Burenkov,
in \textit{VAK-2010 Conf. Proc.} (SAO~RAS, Nizhnii Arkhyz,
2010), p. 127.

10. V. L. Afanasiev and A. V. Moiseev, Astron. Lett. $\textbf{31}$,
194 (2005).

11. A. U. Landolt, Astronom. J. $\textbf{88}$, 439 (1983).

12. S. I. Neizvestny, Izvestiya SAO $\textbf{17}$, 26 (1983).

13. J. Tonry and M. Davis, Astronom. and Astrophys. $\textbf{84}$,
1511 (1979).

14. G. M. Karataeva, I. O. Drozdovsky, V. A. Hagen-
Thorn, et al., Astronom. J. $\textbf{127}$, 789 (2004).

15. V. V. Makarov, V. P. Reshetnikov and V. A. Yakovleva,
Astrofizika $\textbf{30}$, 15 (1989).

16. V. P. Reshetnikov, V. A. Hagen-Thorn, and
V. A. Yakovleva, Astronom. and Astrophys. $\textbf{303}$,
398 (1995).

17. R. I. Jedrzejewsky, Monthly Notices Roy. Astronom.
Soc. $\textbf{226}$, 747 (1987).

18. R. F. Peletier, R. L. Davies, G. D. Illingworth, and
L. E. Davis, Astronom. J. $\textbf{100}$, 1091 (1990).

19. R. Buta, S. Mitra, G. de Vaucouleurs, and H. G. Corwin,
Jr., Astronom. J. $\textbf{107}$, 118 (1994).

20. D. J. Schlegel, D. P. Finkbeiner, and M. Davis, Astrophys.
J. $\textbf{500}$, 525 (1998).

21. S. M. Kent, Astrophys. J. Suppl. $\textbf{59}$, 115 (1985).

22. V. A. Hagen-Thorn, L. V. Shalyapina,
G. M. Karataeva, et al., Astron. Rep. $\textbf{49}$, 958
(2005).

23. V. P. Reshetnikov, Astronom. and Astrophys. $\textbf{416}$,
889 (2004).

24. S. Veilleux and D. E. Osterbrock, Astrophys. J.
Suppl. $\textbf{63}$, 295 (1987).

\end{flushleft}

\end{document}